\documentclass[aps,prb,twocolumn,superscriptaddress]{revtex4}
\usepackage{amsmath,amssymb}
\usepackage{graphicx}
\usepackage{bm}

\newcommand{\cebi} {Ce$_3$Bi$_4$Pt$_3$}
\newcommand{\ubi} {U$_3$Bi$_4$Ni$_3$}

\begin{document}

\preprint{LA-UR-09-00839}

\title{Hybridization driven gap in U$_3$Bi$_4$Ni$_3$: a $^{209}$Bi NMR/NQR study}


\author{S.-H. Baek}
\affiliation{Los Alamos National Laboratory, Los Alamos, NM 87545, USA}
\author{N. J. Curro}
\affiliation{Department of Physics, University of California, Davis, CA 95616, USA}
\author{T. Klimczuk}
\affiliation{Los Alamos National Laboratory, Los Alamos, NM 87545, USA}
\affiliation{Faculty of Applied Physics and Mathematics, Gdansk University of
Technology, Narutowicza 11/12, 80-952 Gdansk, Poland}
\author{H. Sakai}
\affiliation{Los Alamos National Laboratory, Los Alamos, NM 87545, USA}
\affiliation{Advanced Science Research Center, Japan Atomic Energy Agency,
Tokai, Ibaraki 319-1195, Japan}
\author{E. D. Bauer}
\affiliation{Los Alamos National Laboratory, Los Alamos, NM 87545, USA}
\author{F. Ronning}
\affiliation{Los Alamos National Laboratory, Los Alamos, NM 87545, USA}
\author{J. D. Thompson}
\affiliation{Los Alamos National Laboratory, Los Alamos, NM 87545, USA}

\date{\today}

\begin{abstract}
We report $^{209}$Bi NMR and NQR measurements on a single crystal of the Kondo 
insulator \ubi. The $^{209}$Bi nuclear spin-lattice relaxation rate 
($T_1^{-1}$)  shows  activated behavior and is well-fit by a spin gap of 220 K. The $^{209}$Bi Knight shift ($\mathcal{K}$) exhibits
a strong temperature dependence arising from $5f$ electrons, in which 
$\mathcal{K}$ is negative at high temperatures and 
increases as the temperature is lowered.  Below 50 K, $\mathcal{K}$ shows a
broad maximum and decreases slightly upon further cooling.
Our data provide insight into the evolution of the hyperfine fields in a fully gapped Kondo insulator based on $5f$ electron hybridization.
\end{abstract}

\pacs{76.60.-k, 71.27.+a}


\maketitle

\section{Introduction}

Strong electron correlations arising from the interactions between localized
spins and conduction electrons are the origin of the unusual physical
properties of heavy fermion systems.\cite{steglich94}
Among them, Kondo insulators (or heavy fermion semiconductors) are characterized by the emergence of a novel ground state with a small gap
driven by the hybridization between the localized $f$-electrons and the
conduction electrons.\cite{fisk95,fisk96,riseborough00} Typically these 
materials exhibit behavior similar to other heavy fermion metals at high 
temperatures and are well described by the single impurity Kondo model. 
However at low temperature, the semiconducting or insulating behavior 
arises as a result of the periodic array of the local moments, and can be 
described by the periodic Anderson 
model. In a Kondo insulator, the chemical potential lies
within the gap driven by the hybridization, leading to insulating behavior at 
sufficiently low temperatures.  

Whereas there are a number of rare-earth ($4f$) based Kondo insulators, which include
\cebi,\cite{hundley90,severing91} CeRhPn (Pn=As,Sb),\cite{takabatake03}
SmB$_6$,\cite{menth69} and
YbB$_{12},$\cite{kasuya94} actinide-based Kondo insulators are less well studied.
To date, there are reports of  Kondo insulating
behavior in U$_3$Sb$_4$X$_3$ (X=Ni,Pt,Pd)
and U$_2$Ru$_2$Sn.\cite{takabatake90, endstra90, menon98, tran03, rajarajan07}
The recently discovered uranium ternary compound \ubi\ is another potential
candidate \cite{klimczuk08}that might be considered as the uranium counterpart
of the well-known Kondo insulator \cebi. These compounds are not only
isostructural, but also isoelectronic assuming the valance state of U is $>3+$
but $<4+$.
\ubi\ has a body-centered cubic Y$_3$Sb$_4$Au$_3$-type structure with the lattice
parameter $a=9.5793$ \AA~(space group: $I\overline{4}3d$). In this crystal
structure,  the U and Ni atoms occupy unique crystallographic sites with
four-fold inversion symmetry. The single Bi
site [16(c)] has a three-fold axial symmetry
along  the [111] direction. Each Bi atom has three U (and Ni) nearest neighbors
that form an equilateral triangle on a plane perpendicular to its axis of
symmetry.\cite{klimczuk08}
A fully gapped ground state was inferred by specific
heat and photoemission measurements,\cite{klimczuk08} in which specific heat
showed a Sommerfeld coefficient $\gamma_0\sim 0$ and the photoemission
measurements revealed an electronic gap of 840 K.  
However, because Th$_3$Bi$_4$Ni$_3$, which is non-magnetic with Th$^{4+}$, is 
also an insulator, these initial studies were unable to establish the origin of the
gapped state in \ubi.

In this paper, we report $^{209}$Bi NMR and NQR results on a single crystal of
\ubi. The high temperature magnetic susceptibility exhibits Curie-Weiss
behavior, with an effective moment of 3.44 $\mu_B$ consistent with either a
$5f^2$ (U$^{4+}$) or $5f^3$ (U$^{3+}$) electronic configuration at the U site.
The Weiss temperature $\Theta = -117$ K suggests the presence of
antiferromagnetic correlations, yet the ground state is non-magnetic.
In fact, the suppression of $\chi$ from the Curie-Weiss behavior at low 
temperature is consistent with the presence of a hybridization gap. The $^{209}$Bi Knight
shift  $\mathcal{K}$ scales linearly with the magnetic
susceptibility $\chi$ down to a temperature $T^*\sim100$ K.
Below this temperature, we find a Knight shift anomaly which differs from that
observed in the isostructural Ce$_3$Bi$_4$Pt$_3$ Kondo insulator. We also find
the electric field gradient (EFG) as probed by the nuclear quadrupole
frequency $\nu_Q$ exhibits a strong $T$-dependence, which probably
reflects a decrease in the unit cell volume.
The strongest evidence for the presence of a gap is found in the spin-lattice
relaxation rate, $T_1^{-1}$, which exhibits
Arrhenius behavior between 25 K to 150 K. We find an activation energy of 220
K, which we identify with the hybridization gap.

\section{Sample preparation and experimental details}

Single crystals of U$_3$Bi$_4$Ni$_3$ were grown as described in
detail in Ref.~\onlinecite{klimczuk08}. 
$^{209}$Bi (nuclear spin $I=9/2$)
nuclear magnetic resonance (NMR) and nuclear
quadrupole resonance
(NQR) were performed using a conventional phase-coherent pulsed spectrometer.
Knight shift measurements were carried out at the central transition of
the $^{209}$Bi at 29 MHz ($\sim4.2$ T).
The temperature dependence of the nuclear quadrupole frequency ($\nu_Q$) and
the nuclear spin-lattice
relaxation rate, $T_1^{-1}$ were measured in zero field (NQR).  The Bi atoms
are located at axially symmetric sites, and thus the NQR spectrum consists of
four equally spaced transitions.  $T_1^{-1}$ was measured by saturation
recovery at the highest NQR transition, $4\nu_Q = 38.13$ MHz
($\pm\frac{7}{2}\leftrightarrow \pm\frac{9}{2}$).  The relaxation data was then
fit to the following equation appropriate for this transition:
\begin{equation}
\begin{split}
       1-\frac{M(t)}{M(\infty)} &= \frac{16}{715}\exp\left(-\frac{36t}{T_1}\right)
	   + \frac{49}{165}\exp\left(-\frac{21t}{T_1}\right) \\
	   &+ \frac{80}{143}\exp\left(-\frac{10t}{T_1}\right)
	   + \frac{4}{33}\exp\left(-\frac{3t}{T_1}\right). \\
\end{split}
\end{equation}

\section{Results and Discussion}

\subsection{Magnetic Susceptibility and Knight Shift $\mathcal{K}$}

\begin{figure}
\centering
\includegraphics[width=\linewidth]{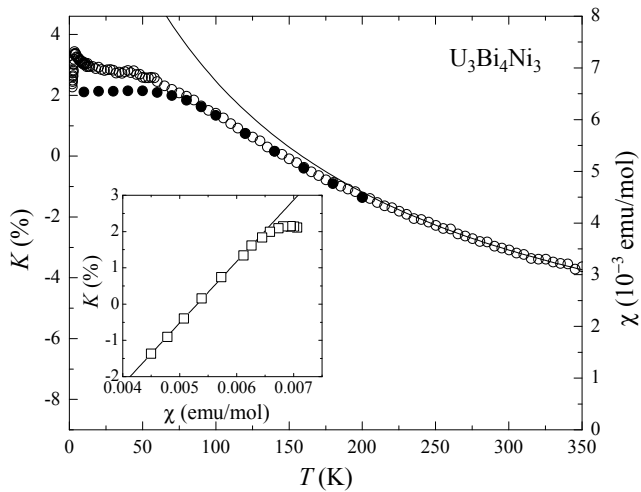}
\caption{\label{fig:K}
Magnetic susceptibility ($\circ$) $^{209}$Bi Knight shift $\mathcal{K}$ ($\bullet$)
as a function of temperature
in a single crystal of \ubi. The susceptibility was measured at 0.1 T. The sharp
decrease of $\chi$ below 3.5 K is attributed to the small amount of
superconducting BiNi on the surface. The solid curve is a Curie-Weiss fit, as
discussed in the text.
The inset shows $\mathcal{K}$ versus $\chi$.
The solid line is a linear fit to the high temperature data, yielding a
hyperfine coupling constant $A_\text{hf}=94(5)$ kOe/$\mu_B$ and
temperature-independent shift $\mathcal{K}_{0}=-9(0.3)$ \%. Below
$T^*\approx 100$ K, the linear relationship breaks down.}
\end{figure}

\begin{figure}
\centering
\includegraphics[width=\linewidth]{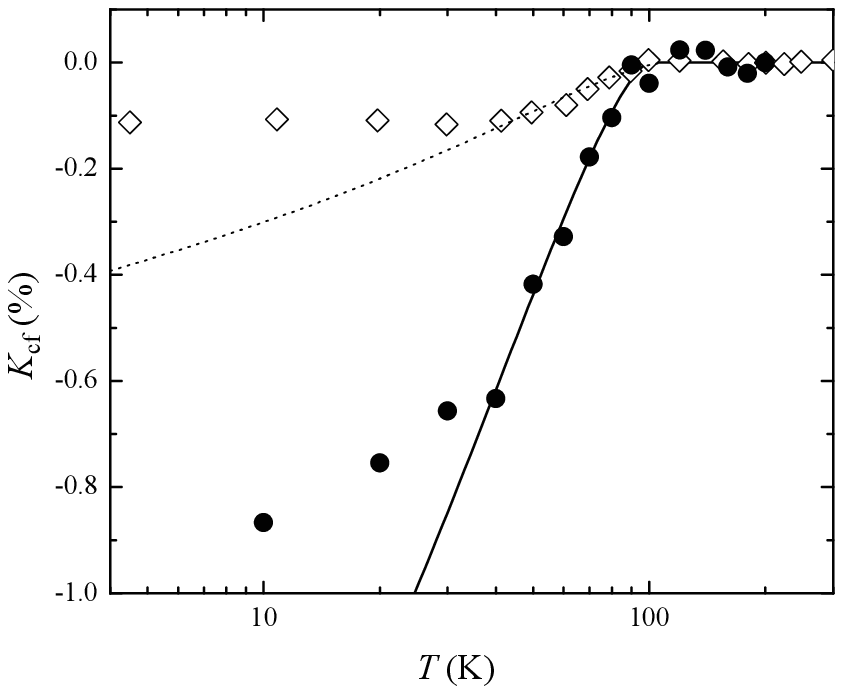}
\caption{\label{fig:Kcf}
$K_{cf}$ versus $T$ in \ubi\ ($\bullet$) and in \cebi\ ($\diamond$).  
The solid and dotted lines are fits as described in the text.}
\end{figure}

Fig.~\ref{fig:K} shows the magnetic susceptibility $\chi$ measured at 0.1 T.
For temperatures above 250 K, a Curie-Weiss fit yields a
magnetic moment of 3.44 $\mu_B$ and a Weiss temperature $\Theta = -117$ K.
As seen in the Figure and the inset, $\mathcal{K}$ tracks the bulk
susceptibility $\chi$ down to approximately 100 K, and below 100 K, the Knight shift passes through a weak maximum.  Plotting $\mathcal{K}$
versus $\chi$,  we obtain the hyperfine coupling
constant $A_\text{hf}=94(5)$ kOe/$\mu_B$ and the temperature-independent shift
$\mathcal{K}_{0}=-9(0.3)$ \%. $\mathcal{K}_{0}$ most likely represents both
orbital and Van-Vleck contributions to the magnetic susceptibility of the U
atoms, whereas the remaining temperature dependent part reflects the
contribution from local $5f$ moments.
As a result of the hybridization between the $5f$ electrons and the itinerant
conduction electrons, the linear  $\mathcal{K}$-$\chi$ relationship breaks
down below $T^*\approx 100$ K, as observed in a number of other heavy fermion
compounds.\cite{curro04}
This Knight shift anomaly arises because there are different hyperfine
couplings to the conduction electron spins, $S_c$ and to the $f$-electron spins,
$S_f$.  Below a characteristic temperature $T^*$, which reflects the scale of
the intersite correlations, the correlation function
$\chi_{cf}=\langle S_c S_f \rangle$ begins to dominate the susceptibility and
gives rise to the difference between $\mathcal{K}$ and $\chi$. In Fig.
\ref{fig:Kcf} we plot the difference,
$\mathcal{K}_{cf} = \mathcal{K} - \mathcal{K}_{0} - A\chi$ versus temperature,
and compare with that observed in \cebi.  The solid and dotted lines are fits to the
two-fluid expression $K_{cf} = K_{cf}^0(1-T/T^*)^{3/2}(1+\log(T^*/T))$ with
$T^*=100(3)$ K for \ubi\ and $T^*=110(5)$ K for \cebi\ that has been used 
successfully in several other heavy fermion 
compounds.\cite{yang08a} In both cases, $|K_{cf}|$ grows with decreasing
temperature below $T^*$, reflecting the growth of correlations, yet below
approximately 40 K, $K_{cf}$ no longer scales with the universal behavior
observed in other heavy fermions.  The origin of this discrepancy is not
known, but we speculate that in the Kondo insulating state there are no
excited heavy quasiparticles and hence $\chi_{cf}$ is modified. The fact that
$\mathcal{K}$ saturates at a finite value at low temperature strongly
suggests that a hyperfine field due to the $5f$ electron spins remains present
in the hybridized, insulating state.

\subsection{Electric Field Gradient}

\begin{figure}
\centering
\includegraphics[width=\linewidth]{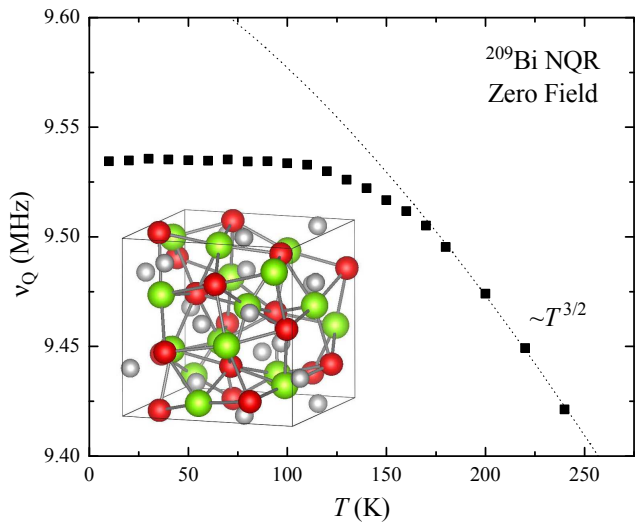}
\caption{\label{fig:nuQ}
$^{209}$Bi nuclear quadrupole frequency $\nu_Q$ as a function of temperature.
Above 200 K, $\nu_Q$ increases with decreasing $T$, as expected from the
phonon contribution $\propto T^{3/2}$ (dotted line). $\nu_Q$ bends over to a
constant value below $\sim 170$ K. INSET: The \ubi\ structure.  The red atoms 
represent U, the green atoms Bi, and the gray atoms Ni.} 
\end{figure}

For nuclei with spin $I>1/2$, the nuclear quadrupole moment, $Q$, interacts 
with the surrounding electric field gradient (EFG) lifting the degeneracy of 
the multiplet in zero field. If the nucleus sits in a site with axial symmetry 
(and lower than cubic) then the quadrupolar Hamiltonian is given by: 
\begin{equation}
\label{eq:nuQ}
       \mathcal{H} = \frac{e^2qQ}{4I(2I-1)h}[3I_z^2-I^2].
\end{equation}
In this case the splitting between adjacent sublevels
$\pm m$ and $\pm (m+1)$ is given by $(m+1/2)\nu_Q$, where $\nu_Q=3e^2qQ/[I(2I-1)h]$ is the
NQR frequency, $m$ is the $I_z$ eigenvalue,
$h$ is Planck's constant, $e$ the electron charge, and $q$ the principal
eigenvalue of the EFG tensor.\cite{slichter}
In \ubi, the $^{209}$Bi nucleus has $I=9/2$ and sits in an axially symmetric
site (16$c$) (see inset Fig. 3). We find three well separated NQR lines at the
frequency of $n\nu_Q$ with $n=2,3,4$ and $\nu_Q=9.531$ MHz at low temperature.
As expected, our results are consistent with axial symmetry (anisotropy parameter $\eta=0$).
Fig.~\ref{fig:nuQ} shows the temperature dependence of $\nu_Q$ measured at the
$4\nu_Q$ transition, in which $\nu_Q$ increases with decreasing temperature and
reaches a plateau below 100 K.

There are two principle contributions to the EFG: the onsite asymmetric
orbitals, and the off-site charges in the lattice.
In most solids, the temperature dependence arises from the lattice vibrations
(phonons)
in which the phenomenological relation $\nu_Q\propto T^{3/2}$ generally
holds.\cite{kaufmann79}
In fact, we find  $\nu_Q\sim T^{3/2}$ down to roughly 170 K, as shown in
Fig.~\ref{fig:nuQ}. However, below this temperature, $\nu_Q$ acquires a
different temperature dependence.  We find that the suppression of $\nu_Q$ is
qualitatively similar to that of the isostructural Ce$_3$Bi$_4$Pt$_3$.
If we adopt the same analysis as made
in Ref.~\onlinecite{reyes94}, the suppression of $\nu_Q$ at low $T$ can be ascribed to
the change of the lattice parameter.  In this case, $\nu_Q$ may be written as
the following form:
\begin{equation}
\label{eq:nuQ2}
          \nu_Q= \nu_Q^0 - a/V(T) - bT^{3/2},
\end{equation}
where $V$ is the cell volume, $\nu_Q^0$ the temperature independent term, and
$a,b$ the positive constants.  In this case, our data suggest that below $170$
K, the cell volume $V(T)$  decreases with
decreasing temperature. Although there is no available
data of the lattice parameter at low temperatures yet,
we suggest that this apparent decrease of the cell volume is related to the
onset of hybridization of $5f$ electrons with the conduction band.  This onset
correlates well with our Knight shift measurements showing $T^*\sim 100$ K.

\subsection{Nuclear spin-lattice relaxation rate}

\begin{figure}
\centering
\includegraphics[width=\linewidth]{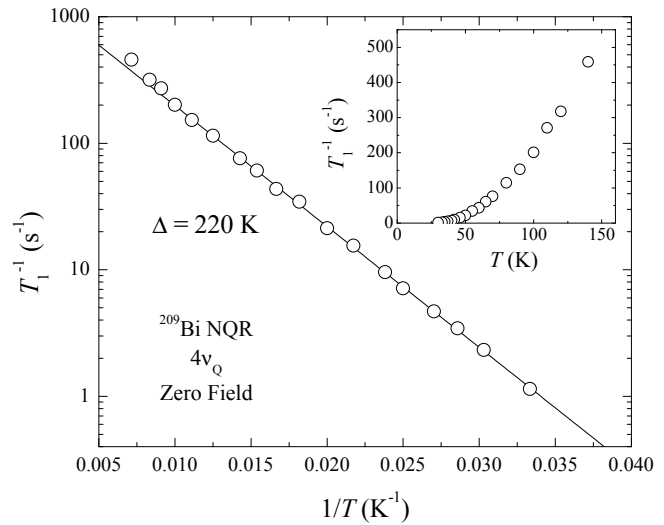}
\caption{\label{fig:T1}
$T_1^{-1}$ of $^{209}$Bi NQR in zero field measured at $4\nu_Q$ line.
$T_1^{-1}$ versus $1/T$ plot reveals the activation behavior with the
single activation energy $\Delta=220$ K. Inset shows
that $T_1^{-1}$ decreases exponentially with decreasing temperature.   }
\end{figure}

The nuclear spin-lattice relaxation rate $T_1^{-1}$  is shown as a function of
$T$ in Fig. \ref{fig:T1}. $T_1^{-1}$ drops dramatically between 150 K
and 25 K, and is well fit by an Arrhenius expression:
$T_1^{-1} = A \exp(-\Delta/T)$, with $\Delta = 220(10)$ K.  This result reveals
the gapping of excited quasiparticles at low temperature, in agreement with
resistivity measurements that show this material to be non-metallic, and
specific heat measurements that reveal a vanishing Sommerfeld coefficient.\cite{klimczuk08} We would not expect these behaviors to arise from only depopulation of an excited crystal field level, such as found in U$_3$Sb$_4$Ni$_3$.\cite{rainford96}
Furthermore, the high quality of the Arrhenius fit rules out the possibility
of an impurity band near the chemical potential. Photoemission results have
revealed a gap in the density of states of 840 K,\cite{klimczuk08} roughly four times the gap
measured by NMR.  The origin of this difference is not clear, but may be
related to spin versus charge excitations. Similar results were found in
\cebi\ (ref.~\onlinecite{reyes94}), and we argue that this behavior is a signature of a Kondo
insulating ground state in \ubi. It is curious that the Knight shift reveals a
large hyperfine field at low temperatures, whereas $T_1^{-1}$ indicates the
absence of any substantial fluctuating hyperfine fields.  \textit{A priori},
these results appear somewhat contradictory, as the same hyperfine field is
responsible for both phenomena. However, if the electronic system is
described by a periodic Anderson lattice with two dispersing bands of partial
$f$-character and a chemical potential lying in the  hybridization gap, then at
low temperature it is possible that the occupied lower band may consist of a
significant $f$-character that will contribute to a static magnetization.
Scattering of heavy mass quasiparticles, on the other hand, will not be
present for temperatures much less than the gap value.

We note that this gap value is smaller than the value reported
in ref.~\onlinecite{klimczuk08} but still larger than the gap values of most
Ce-based Kondo insulators.\cite{riseborough00}  
This difference between the $4f$ and $5f$ gap values
may indicate the better hybridization
between U $5f$ electrons and conduction electrons from Bi $6p$ and/or Ni $3d$ 
bands in \ubi.  We note that $T^*/\Delta\sim 0.48$ for \ubi, 
whereas $T^*/\Delta\sim 0.61$ for \cebi.  
The origin of this discrepancy is unclear, but suggests that the energy scales 
determining $\Delta$ ($T_K$) and $T^*$ differ, as argued recently 
in ref.~\onlinecite{yang08a}.

\section{Conclusion}

We presented $^{209}$Bi NMR/NQR measurements on a single crystal  of \ubi.
Knight shift and susceptibility measurements reveal a Knight shift
anomaly that probably arises due to different hyperfine couplings to the $5f$
electron spins and to the itinerant conduction electrons.  The Knight shift
anomaly is in reasonable agreement with the two-fluid model, and is similar with
observations in \cebi.\cite{reyes94}  The temperature dependence of the EFG suggests a
reduction of the unit cell volume at low temperatures,  and may be related to
the onset of hybridization between the lattice of nearly localized $5f$ electrons and the
conduction band.
The most striking result is the spin lattice relaxation rate, which exhibits 
activated behavior between 150 K and
25 K, with an activation energy $\Delta=220$ K. Our results suggest that \ubi\
is a $5f$-based Kondo insulator, and that strong correlations play an important
role in this structural class of U based compounds.  Detailed electronic
structure calculations should help to shed light on the role of correlations
in these materials. Finally, our results provide new constraints on the
interpretation of NMR in $f$-electron systems in the limit of vanishing
quasiparticles.  

\section*{Acknowledgments}

We thank H. Lee, S. Savrasov, Y.-F. Yang, and T. Durakiewicz for 
discussions. Work at Los Alamos National Laboratory was performed under the
auspices of the US Department of Energy, Office of Science.

\bibliography{/mydocuments/mydata/mybib}

\end{document}